\documentclass[fleqn,usenatbib]{mnras}

\usepackage{newtxtext,newtxmath}

\usepackage[T1]{fontenc}
\usepackage{subfigure}
\usepackage[justification=centering]{caption}

\DeclareRobustCommand{\VAN}[3]{#2}
\let\VANthebibliography\thebibliography
\def\thebibliography{\DeclareRobustCommand{\VAN}[3]{##3}\VANthebibliography}

\usepackage{graphicx}	
\usepackage{amsmath}	

\usepackage{threeparttable}

\title[Time-domain analysis of flares from AD Leo]{Time-domain analysis of multi-waveband flares from AD Leonis}

\author[Zhang et al.]{
Xuying Zhang,$^{1}$
Jixuan Li,$^{1}$
Yang Gao$^{1,2}$\thanks{
  E-mail: gaoyang25@mail.sysu.edu.cn}
and Qian Lei$^{2,3}$
\\
$^{1}$School of Physics and Astronomy, Sun Yat-Sen University, Zhuhai 519082, Guangdong, People's Republic of China\\
$^{2}$CAS Key Laboratory of FAST, National Astronomical Observatories, Chinese Academy of Sciences, Beijing 100101, People's Republic of China\\
$^{3}$University of Chinese Academy of Sciences, Beijing 10049, People's Republic of China
}

\date{Accepted 2024 April 23. Received 2024 April 22; in original form 2023 September 4}

\pubyear{2023}
\graphicspath{{./}{fig/}}
\begin{document}
\label{firstpage}
\pagerange{\pageref{firstpage}--\pageref{lastpage}}
\maketitle

\begin{abstract}
Radio bursts of magnetically active stars reveal the intensity and activity of the stellar magnetic field. 
They may also be related to the planets around the stars. 
We monitored a radio-active star, AD Leonis, 3000 seconds per day for 17 days in November 2020, and 5000 seconds per day for 5 days in July 2023 with the Five-hundred-meter Aperture Spherical radio Telescope (FAST). 
Based on the simultaneous flux increases in Stokes I and Stokes V, one left-hand circular polarized radio burst is identified. 
The $\sim50\%$ degree of circular polarization indicates the burst being originated from non-thermal radiation related to the stellar magnetic field. 
Combining the newly discovered burst with previous observations of radio and X-ray bursts from AD Leonis, we did a periodicity analysis for the 49 bursts in total.
No periodicity with confidence level $>3\sigma$ is found, 
while a candidate period of 3.04 days at $\approx 2\sigma$ confidence level is presented and discussed.
Results of recent FAST observations and the periodicity analysis suggest a more compact campaign of observation toward this source, from which a more optimistic result of period search could be achieved.
\end{abstract}




\section{Introduction}

Flares are sudden, violent ejections of energetic particles caused by magnetic reconnection (MRX) in stellar atmospheres \citep{benz2010}. Stellar flares are observed as intense bursts of radiation in radio to UV and X-ray wavebands. 
For magnetically active stars such as M dwarfs, the flux of their radio bursts can be $10^3$ higher than solar radio bursts \citep{dulk1985}. 
Systematic radio observation campaigns towards flare stars have been carried out with large telescopes or telescope arrays such as Arecibo \citep[eg. ][]{abada-simon1994,smith2005}, VLA \citep[eg. ][]{jackson1989,villadsen2019} and LOFAR \citep[eg. ][]{callingham2021} in the past decades.
Flare radiation mechanisms are studied based on the observations. 
Due to the high brightness temperature and circular polarization, the mechanisms discussed usually focus on coherent emissions, i.e. electron cyclotron maser \citep[ECM, ][]{melrose1982} and plasma emission, as well as the possible synchrotron emission. 

For stars with a planet enclosed in the stellar magnetosphere, the magnetic interaction between the star and the planet \citep[SPI,][]{cuntz2000} may also lead to energetic flares from the system. 
Although radio emission from exoplanets has been the target of radio astronomy for many decades since \citet{yantis1977}, no radio signal from an exoplanet has been definitely recognized \citep[cf.][]{turner2021}.

However, observations of the Jovian satellite system, in which emissions from Jupiter-satellite interactions are recognized \citep{zarka2018}, reveal the detectability and periodicity of planet-satellite interactions. Considering similar interaction mechanisms between stars and planets,
more efforts are paid to identifying the signals of exoplanetary SPIs \citep{vedantham2020, callingham2021, pope2021, pineda2023, route2023}, and periodicity analysis on bursts and flux variations from star-planet systems is widely adopted \citep{perez-torres2021, trigilio2023, loyd2023, acharya2023,bloot2023}. 

As summarized in \citet[][]{zarka2007}, there are 3 types of SPIs: (i) Unipolar interaction of a planet with the stellar magnetosphere which generates Alfv{\'e}n waves, similar to the Io-Jupiter system \citep{saur2013}; (ii) Dipolar interaction caused by the MRX between the star and magnetized planet, similar to the Ganymede-Jupiter system \citep{zarka2018}; (iii) Magnetospheric interaction caused by a developed exoplanetary magnetosphere interacting with the stellar wind, which is a counterpart of the interaction between the Earth's magnetosphere and the solar wind. 
As a result, constraints of an exoplanetary system, especially on its magnetic field, may be inferred from SPI. 
However, confirming an SPI signal in observations is complicated because of the difficulty in ruling out stellar activities that are irrelevant to SPI \citep{pineda2023}. 
Due to the non-axisymmetry of the stellar magnetic field and the change of the observer’s viewing angle, SPI-induced flares may be modulated by periods of either the planetary orbit or the beat of a planetary orbit and stellar rotation \citep{gao2021}. 
Statistically, periodicity analysis based on an extensive dataset of flares is capable of revealing the existence of planetary orbital modulation. 

One of the fundamental ways of searching for periodicity in astronomical observations is phase-folding \citep{vanderplas2018}. 
This process involves introducing a test period, folding the observed time series to a function of phase, and then calculating well-defined statistics of the phased data to assess the presence of periodicity. 
Phase-folding methods such as the minimum string length \citep{dworetsky1983} and quadratic mutual information \citep{huijse2018} have been utilized to search for periods of stellar radio emissions \citep{perez-torres2021} and Fast Radio Bursts (FRBs) \citep{pastor-marazuela2021}, concentrating on the continuous change of the observable with the period as the only parameter in the model. 
Targetting the period of flares' occurrence as the primary focus and introducing the width of the active phase as another parameter, the Phase-folding probability Binomial Analysis (PBA) has been proposed for periodicity analysis of bursty signals \citep[][]{gao2021,li2024}. 

The dMe star AD Leonis is one of the most extensively observed radio flare stars, with flares identified in different wavebands, i.e., radio \citep{lang1983, lang1986,jackson1989, gudel1989, bastian1990, abada-simon1994, stepanov2001,zaitsev2004, osten2006, osten2008, villadsen2019, zhang2023, mohan2024}, optical \citep{pettersen1984,hawley1991,hawley2003,crespo-chacon2006,dimaio2020,muheki2020,bai2023,wollmann2023}, EUV \citep{hawley1995,cully1997,gudel2003,sanz-forcada2002}, and X-ray \citep{sciortino1999,favata2000,vandenbesselaar2003,robrade2005,namekata2020,stelzer2022}.
We use the Five-hundred-meter Aperture Spherical radio Telescope \citep[FAST,][]{nan2006a, nan2011} to monitor its radioactivity at $L$-band, expecting for detection of radio bursts with flux density higher than the systematic error of a few mJy of the main beam. 
Another reason to choose AD Leonis is that it has a low spin axis inclination of $\sim$15 degrees \citep{tuomi2018}, with its axis nearly pointing toward the observer. 
This `pole-on' geometry indicates that detecting a planet around AD Leo using the conventional radial velocity method could be challenging. 
Consequently, the ongoing debates regarding the existence of a planet orbiting AD Leo \citep{tuomi2018, carleo2020, kossakowski2022, carmona2023} might be checked through time-domain analysis of its radio flares.
Because the heating of stellar chromosphere by energetic particles from SPI or stellar flares also induces an increase in thermal X-ray emission \citep{smith2005, benz2010}, we enlarge the dataset to include both radio and X-ray flares to search for potential periods.
  
Observation and data processing are specified in Section \ref{sec:observation}, with 
  one flare identified in Section \ref{sec:flares}.
Periodicity analysis is made in Section \ref{sec:period}.
The conclusion is made in Section \ref{sec:conclusion}.

\section{Observations and Data Processing}
\label{sec:observation}
The target AD Leo with right ascension $\alpha$(J2000)=10$^{h}$19$^{m}$36$^{s}$ and declination $\delta$(J2000)=19$^{\circ}$52$^{\prime}$12$^{\prime\prime}$ is a M3.5V star 
at the age of 25 - 300 Myr  \citep{shkolnik2009}.
It is located 4.9 pc from us, possesses a mass of $0.42M_{\odot}$ \citep{morin2008}, and has a spin period of 2.23 days \citep{shkolnik2009,fouque2023}.
Within the main beam of FAST when it directs toward AD Leo, there is a bright extra-galactic radio source NVSS J101945+195123 with flux $\sim$250 mJy at 1.4 GHz. Previous observations did not show time variation of this source \citep{condon1998, vanvelzen2015}.

More than ten radio flares from AD Leo have been observed previously, with research focused on single burst events. 
To enrich the dataset of flares for periodicity analysis, we conducted a monitoring campaign of AD Leo using FAST from 8 to 24 in November 2020; from 25 to 28, and on 31 in July 2023.  
Observations in 2020 cover seven stellar spin periods, with 3000 seconds of observation time spent every day. 
The observations employ the tracking mode with spectral backend to record four polarizations, using 1 second sampling time and 400 MHz bandwidth centred at 1250 MHz. 
Observations in 2023 cover more than two periods of stellar spin, with 5000 seconds of monitoring time each day. 
In order to record baseline temporal variations, we use the phase-referencing mode (on-off mode) in 2023, i.e., after tracking AD Leo for every $\approx$290 seconds, the main beam is switched to $\alpha$(J2000)=10$^{h}$19$^{m}$59$^{s}$ and $\delta$(J2000)=19$^{\circ}$52$^{\prime}$12$^{\prime\prime}$ to record the off-source flux for $\approx$10 seconds.
The spectral backend with 0.1 second sampling time and pulsar backend with 1 millisecond sampling time are adopted simultaneously on the 19-beam receiver. 
The spectrum band setup remains the same with 2020 observations.

\subsection{Calibration}
The FAST 19-beam receiver array contains a temperature-stabilized noise injection system, which can be used for calibrating polarization and flux density. 
During the 2020 observations, however, the reference noise injection period (i.e. the duration between the start time of two adjacent successive `on' states of the noise source) was set shorter than the sampling period. 
Consequently, we calibrate the flux density using neutral hydrogen lines instead. For polarization calibration, we adopt the noise diode data from other observations during the period we observe AD Leo to do the calibration. 
For the 2023 observations, conventional noise diode calibrations are adopted for both flux density and polarization.

\subsubsection{Flux Calibration}
For 2023 observations, the high noise diode of $\approx$12 K is used for flux calibration.
We first find the noise diode `on' data and subtract the adjacent noise diode `off' counts from it to get the pure noise diode counts.
The ratio between the noise diode temperature and the counts, after divided by the gain of FAST beam M01 \citep[Table 5 of][]{jiang2020}, is the calibration coefficient for flux density.

For 2020 observations, flux calibration is completed by using the galactic HI lines observed in 2020 and 2023 at the same point of the sky, with its flux density assumed constant over the years. 
We first calculate the FWHMs of HI lines, then average their antenna temperature in 2023 and their counts in 2020 within the FWHM respectively. 
The ratio between the mean antenna temperature in 2023 (identical to the 2020 value) and the mean counts in 2020, divided by the gain of FAST, is the calibration coefficient of flux density.  
In the 2020 campaign, HI lines observed in 13 days out of the 17 days overlapped with other signals (cf. Appendix \ref{appendix:removal}). Hence, we use the 4 `clean' HI lines observed on November 15, 18, 21, and 24 for calibration, with the calibration coefficient being the average of them.


\subsubsection{Polarization Calibration}
\label{subsubsec:pc} 
To record and characterize the signal polarization effectively, FAST employs a pair of orthogonal linear polarization probes. 
These probes are essential for recording both the intensity and phase of the signals in two orthogonal directions, denoted as $x$ and $y$. 
The system generates four cross-products from the two orthogonal outputs $e_x$ and $e_y$, namely 
\begin{eqnarray}
AA=\langle e_x e_x^*\rangle, \nonumber \\
BB=\langle e_y e_y^*\rangle, \nonumber \\
CR=\langle {\rm Re}[e_x e_y^*]\rangle, \nonumber \\
CI=\langle {\rm Im}[e_x e_y^*]\rangle. \nonumber
\end{eqnarray}
The 4 Stokes parameters can be achieved from the cross-products through $I=AA+BB$, $Q=AA-BB$, $U=2\times CR$, $V=2\times CI$. 
The phase delay $\delta$ between the 2 orthogonal signals is calculated by 
\begin{eqnarray}
\tan{\delta} = CI / CR.  \nonumber  
\end{eqnarray}
The injection from the noise diode serves as the reference signal, which satisfies $AA_{\rm ref} = BB_{\rm ref}$, $\delta_{\rm ref} = 0$. 
Therefore, any observed change of $AA_{\rm ref} / BB_{\rm ref}$ and $\delta_{\rm ref}$ at the backend represents the impact of the receiver on the signals that enter the system. So for the signal from the target, $\Delta (AA_{\rm sig} / BB_{\rm sig}) = \Delta (AA_{\rm ref} / BB_{\rm ref})$, $\Delta\delta_{\rm sig} = \Delta\delta_{\rm ref}$.
After removing these impacts from the observation data of astronomical sources, one finishes the polarization calibration.

For 2023 observations, the above calibration is directly applied. 
Polarization calibration for 2020 observations is achieved by using the noise diode data from other observations with the same mode and backend settings during the period we observe AD Leo. 
$\Delta (AA_{\rm ref} / BB_{\rm ref})$ in these observations can be consider constant with a mean of $0.12\pm0.16$, which is used in the calibration.
Conversely, $\Delta\delta_{\rm sig}$ correlates with environmental temperature and we calculate it for every second to correct the phase delay in the observation signals.
More details on the 2020 polarization calibration can be found in Appendix \ref{appendix:polarization calibration}.

\subsection{Noise and Background Flux Removal}
\subsubsection{RFI and Standing Waves}
As shown in the bottom panel of Figure \ref{fig:magflare} (a), we first plot the light curves of the mean continuum flux density at HI-protected waveband 1410-1430 MHz.
At the full waveband of 1050-1450 MHz, the radio frequency interference (RFIs) are removed and the dynamic spectra are presented. 
The wavebands of 1125-1300 MHz in 2020 observations and 1150-1300 MHz in 2023 observations are firstly masked as they were heavily impacted by RFI. 
For the rest wavebands, the Asymmetrically reweighted Penalized Least Squares \citep[ArPLS, ][]{baek2015} method is adopted to the mean spectra calculated every 10 seconds to remove RFIs \citep{zeng2020, zhang2022}. 
Standing waves and other interferences are also removed from the continuum by iteratively using ArPLS.
Finally, we mask discrete frequencies with $>$3$\sigma$ flux increases in the same mean spectra. 
The above removals are carried out for both Stokes I and V.

The standing waves become more obvious in the HI-protected waveband due to the absence of RFI (Figure \ref{arpls}). They are periodic fluctuations arising from reflections between the FAST's main reflector and the steel cables of the receiver cabin \citep{jiang2020}. 
For standing waves, we smooth the negative peaks in the SED before applying the ArPLS method (cf. Figure \ref{arpls} (c)).

\subsubsection{Background Flux}
The background noise $T_{\rm sys}$ is composed of the systematic noise of the receiver ($T_{\rm rec}$), continuum brightness temperature of the sky ($T_{\rm sky}$), emission from the Earth's atmosphere ($T_{\rm atm}$) and radiation from the surrounding terrain ($T_{\rm scat}$), i.e. \citep{jiang2020},
\begin{equation}
    T_{\rm sys} = T_{\rm rec} + T_{\rm sky} + T_{\rm atm} + T_{\rm scat}.
\end{equation}

In 2023, the background flux has been tracked by off-source observations in the phase-referencing mode. 
We adopt interpolation to form the off-source light curve and subtract it from the on-source one to get the light curve of the source. 
For the source NVSS J101945+195123 included in beam M01, we convolve its flux with the beam shape and then subtract it from the source flux.
Finally, the flux from AD Leo is achieved and shown in Figure \ref{fig:magflare}. 

There is no off-source observation in 2020, so the background flux is estimated each day by roughly averaging the flux over relatively `quiet' periods in Stokes I. This is done for each 1 MHz bandwidth. 

\subsection{Error Analysis}
Two dominant sources of flux variations are the variation in the background flux $T_{\rm sys}$, and the fluctuation of calibration coefficient. 
The background flux variation is mainly caused by different zenith angles (ZA) of the telescope, so it is a long-term variation.
Its typical value is $\sim$0.4 K, corresponding to $\sim$25 mJy \citep{jiang2020} in both 2020 and 2023 observations. 
In 2020, the injected noise was also part of the background, which leads to a random fluctuation of $\sim 1\% \times 12$ = 0.1 K, corresponding to $\sim$6 mJy.
The above variations/fluctuations are additives to the flux density. 

The other source of error is the calibration coefficient fluctuation. 
Firstly, the $\sim$1\% random error of telescope gain \citep[][]{jiang2020} should be counted in for all observations. 
Observations in 2023, which utilize the noise diode for calibration, have an additional calibration uncertainty of $1\%$ originating from the noise diode temperature fluctuation. 
For 2020 observations, HI lines are used for calibration. 
So the total error of 2023 observations ($\sim$2\%) should be adopted as this additional calibration uncertainty. 
The total relative error of Stoke I is $\sim$3\% in 2020 and $\sim$2\% in 2023.

For Stokes V, the error of $\Delta (AA_{\rm ref} / BB_{\rm ref})$ and the phase delay $\Delta\delta_{\rm ref}$ are also considered (cf. Section 2.1.2). 
However, the total error is small due to the $\sim$0 K background flux.

\section{Flare Identification}
\label{sec:flares}

\begin{figure*}
    \includegraphics[width=\linewidth]{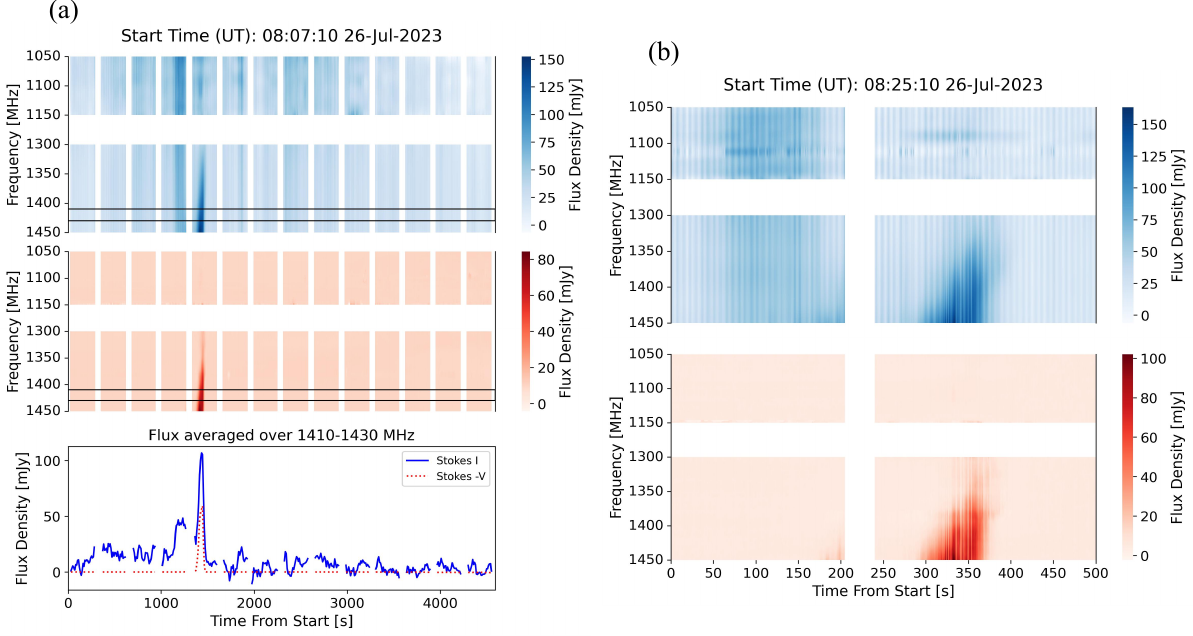}
    \caption{Non-thermal flare identified on July 26, 2023, with 10 second time resolution (a), and 1 second time resolution (b). For panels (a) and (b), dynamic spectra of Stokes I and V are shown in the blue and red sub-panels, respectively. White areas are masked due to RFI and off-source observations. In panel (a), the light curves of continuum flux averaged over 1410-1430 MHz (framed with black rectangles in the dynamic spectra) are shown in the bottom panel, with blue and red lines for Stokes I and V, respectively. In panel (b), there are vertical white bands during the burst period, the origin of which is uncertain.}
    \label{fig:magflare}
\end{figure*}

\begin{figure*}
    \includegraphics[width=\linewidth]{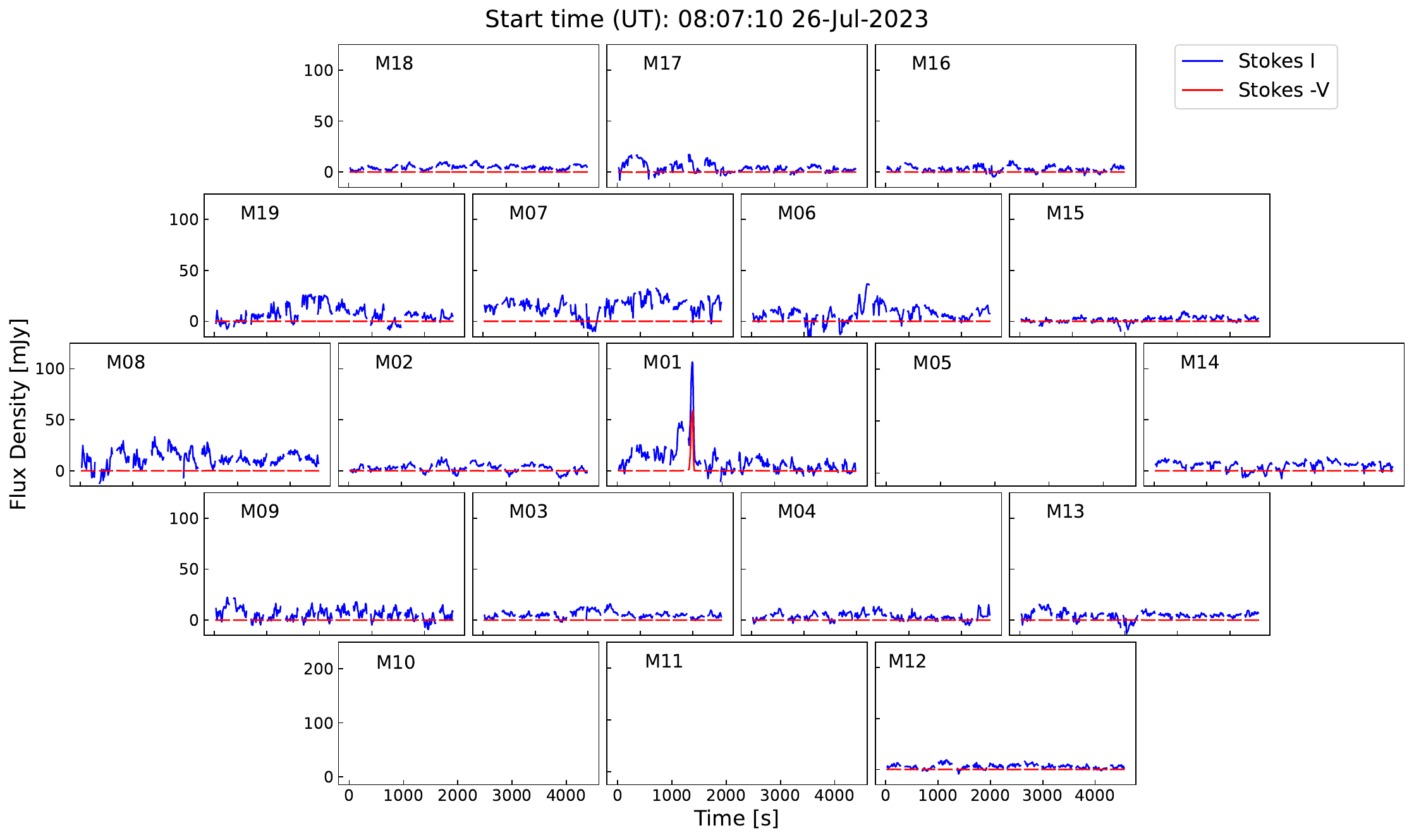}
    \caption{Light curves of continuum flux density averaged over 1410-1430 MHz for Stokes I (blue lines) and Stokes V (red lines) on FAST 19-beam receiver. For the beams in the blank: there are bright radio sources in the off-source observation of M05, and the data recorded for M10 and M11 are unstable or halted.}
    \label{fig:19beams}
\end{figure*}

Dynamic spectra of the total intensity Stokes I and circular polarization Stokes V, as well as the corresponding light curves, indicate whether a flare occurred. 
Thermal fluctuations from either the source or the background are not of interest here. The crucial criterion of a non-thermal radio flare is the simultaneous increase in Stokes I and V. Such flare, if identified, is more likely to originate from the magnetic field of AD Leo.  

\subsection{Flare Identified in FAST Observation}
\label{subsubsec:Flares} 

A candidate flare with a sharp increase in Stokes I was found on July 26, 2023. According to Figure \ref{fig:magflare}, the flux increase in Stokes I lasts $\sim$100s and peaks at $\sim$100 mJy. 
There is a synchronous increase in Stokes V, with the left-hand circular polarized flux reaching $\sim$50 mJy. 
It is also seen from Figure \ref{fig:19beams} that such variations are not detected in other beams of the FAST 19-beam receiver, which means the increase of flux in the main beam (M01) should not be RFI or background fluctuations.
The $\sim$50\% circular polarization indicates that the signal may originate from the magnetic field of the star.

No candidate flare is found on other days of the 2023 campaign as well as in the 2020 campaign.
There are long-term variations of Stokes I in some observations in the two campaigns, but they are not considered as flares as no synchronous variation in Stokes V is detected.


As shown in Figure \ref{fig:magflare} (b), the candidate flare appears in 1300-1450 MHz, with clear frequency drift and $\sim$50\% left circular polarization. 
In Table \ref{dataset1} we summarize the radio and X-ray flares from the literature. 
On the one hand, some observations identified 100\% left-hand circular polarized flares at millisecond resolution. 
The <100\% circular polarization of the flare in Figure \ref{fig:magflare} may be due to the low time resolution. 
Each time step of 1 second may contain a number of flares with 100\% circular polarization, and the lags between them should have zero or low circular polarization as inferred from the light curve. So the apparent polarization is lower than 100\%. 
On the other hand, non-100\% circular polarized short flares were also observed previously which can be low polarization synchrotron emissions mixed with 100\% polarized ECMs \citep{smith2005}. 
We will look into more detailed structures of this flare in the FAST pulsar-end data and compare it with multi-waveband observations in follow-up work. 

\subsection{Possible Origin of the Flare}
Under the assumption that the spatial scale of the emission source is the same as the stellar radius \citep[$\sim$0.436 $R_{\odot}$, ][]{houdebine2016}, we can derive the lower limit of the brightness temperature for this flare to be $\sim10^{11}$ K using Equation (6) in \cite{zic2019}. 
The high brightness temperature and degree of circular polarization indicate an origin of non-thermal emission. 
Among the plausible mechanisms, ECM and plasma emission are considered \citep{zic2019,zic2020,vedantham2021,mohan2024}. \citet{zhang2023} have shown that plasma emission is less expected in the FAST $L$-band, we look more into the ECM mechanism.

ECMs are proven to possibly occur in the stellar magnetosphere and produce significant radio emission \citep{melrose1982}. This coherent mechanism yields high circular polarized (sometimes reaching 100\%) and broadband non-thermal radiation \citep{dulk1985,treumann2006}.
ECM has an upper-frequency limit (i.e. cut-off frequency) that is proportional to the strength of the magnetic field of the source object. 
At frequencies higher than the ECM upper limit, broadband synchrotron emission from relativistic electrons could be dominant, 
if there are strong stellar flares or other bursty processes that produce large amounts of energetic electrons.

AD Leo has a `pole-on' geometry and a dipole-dominated axisymmetric magnetic field \citep{morin2008,lavail2018,bellotti2023}.
Therefore, the magnetic fields at the poles of the star are parallel to the line of sight (LOS) and gradually change to directions perpendicular to the LOS as the stellar latitude decreases. 
Then if we consider an ECM origin of this left-hand circular polarized radio flare from the stellar dipole field, AD Leo should have its south magnetic pole pointing toward us \citep[cf.][]{villadsen2019}.
However, it is also possible that the radio emission comes from coronal loops of the star, which may explain why both left and right circular polarized radio flares have been observed before.
The strength of the magnetic field can be derived using the cut-off frequency in ECMs. 
As the cut-off frequency should be higher than the highest frequency 1450 MHz observed here, the maximum magnetic field strength of AD Leo should be >500 Gauss \citep[calculated using equation (6) in ][]{farrell1999}.
This is consistent with the recent results from Zeeman-Doppler Imaging \citep{reiners2022,cristofari2023,bellotti2023}. 
The frequency ranges of flares detected in this paper and in other observations \citep[eg.][]{villadsen2019,zhang2023} may also imply the strength of the magnetic field in corresponding coronal loops \citep[cf. discussions in][]{lynch2015}. 

\section{Periodicity Analysis}
\label{sec:period}

\subsection{Data Collection}

\begin{table*}
  \caption{AD Leo flares in Dataset I.}
  \begin{threeparttable} 
\begin{tabular}{cccclcccc}
\hline
ID & MJD & Telescope                                                                      & Reference                                                               &  & ID & MJD & Telescope  & Reference   \\ \hline
1  & 44372.510    & Einstein IPC              & \citet{favata2000}                   &  & 26 & 58568.509    & NICER      & \citet{namekata2020}    \\
2\tnote{h}  & 45366.233  & Arecibo     & \citet{lang1983}                    &  & 27 & 58585.650    & NICER      & \citet{namekata2020}     \\
3  & 46146.264   & VLA                                                                       & \citet{jackson1989}                                                                &  & 28 & 58586.503   & NICER      & \citet{namekata2020}                                 \\
4  & 46261.763   & Arecibo                                                                        & \citet{lang1986}                                                             &  & 29 & 58823.380   & NICER      & NICER Archive Website\textsuperscript{N}     \\
5\tnote{h}  & 47103.460   & Arecibo                                                               & \citet{bastian1990}                                                               &  & 30 & 58824.337    & NICER      & NICER Archive Website\textsuperscript{N}     \\
6  & 47106.455   & \begin{tabular}[c]{@{}c@{}}Effelsberg, Jodrell\\ Bank and Arecibo\end{tabular}   & \citet{gudel1989}                                                         &  & 31 & 58826.886   & NICER      & NICER Archive Website\textsuperscript{N}     \\
7  & 48238.403    & Arecibo                                                                        & \citet{abada-simon1994}                                                         &  & 32 & 58828.335   & NICER      & NICER Archive Website\textsuperscript{N}     \\
8  & 48384.606     & PSPC                                                                          & \citet{favata2000}                                                              &  & 33\tnote{h} & 58830.031    & NICER      & NICER Archive Website\textsuperscript{N}     \\
9  & 48385.103     & PSPC                                                                    & \citet{favata2000}                                                            &  & 34 & 58831.609    & NICER      & NICER Archive Website\textsuperscript{N}     \\
10 & 49031.176    & Arecibo                                                                      & \citet{abada-simon1994}                                                          &  & 35 & 58832.460    & NICER      & NICER Archive Website\textsuperscript{N}     \\
11 & 50206.992    & ASCA                                                                      & \citet{favata2000}                                                                &  & 36 & 58832.778    & NICER     & NICER Archive Website\textsuperscript{N} \\
12 & 50207.224    & ASCA                                                                       & \citet{favata2000}                                                                &  & 37 & 58836.137    & NICER     & NICER Archive Website\textsuperscript{N}  \\
13 & 50207.305    & ASCA                                                                          &\citet{favata2000}                                      
                        &  & 38 & 58836.367    & NICER       & NICER Archive Website\textsuperscript{N} \\
14\tnote{h} & 50587.790    & Effelsberg                  & \citet{stepanov2001}                                                                             
                     &  & 39 & 58838.822    & NICER       & NICER Archive Website\textsuperscript{N}  \\
15 & 51300.718    & BeppoSAX                                                                       & \citet{osten2006}                                                               &  & 40 & 59537.242    & XMM-Newton       & \citet{stelzer2022}  \\
16 & 51306.475    & BeppoSAX                                                                       & \citet{gudel2003}                                                               &  & 41 & 59550.865    & FAST & \citet{zhang2023} \\
17 & 51313.170    & BeppoSAX                                                                        & \citet{gudel2003}                                                               &  & 42\tnote{h} & 59551.884    & \begin{tabular}{@{}c@{}} FAST \\ and uGMRT\end{tabular}       & \begin{tabular}{@{}c@{}} \citet{zhang2023}\\ \citet{mohan2024}\end{tabular}   \\
18 & 52043.867    & \begin{tabular}[c]{@{}c@{}}XMM-Newton\\ and VLA\end{tabular}                & \begin{tabular}[c]{@{}c@{}}\citet{robrade2005};\\ \citet{smith2005}\end{tabular}          &  & 43 & 59616.561    & NICER       & NICER Archive Website\textsuperscript{N}   \\
19 & 52044.146    & \begin{tabular}[c]{@{}c@{}}XMM-Newton\\ and VLA\end{tabular}                & \begin{tabular}[c]{@{}c@{}}\citet{robrade2005};\\ \citet{smith2005}\end{tabular}          &  & 44 & 59617.593    & NICER      & NICER Archive Website\textsuperscript{N}     \\
20 & 52803.941    & Arecibo                                                                            & \citet{osten2006}                                                  &  & 45 & 59628.193    & NICER      & NICER Archive Website\textsuperscript{N}     \\
21 & 52804.837    & Arecibo                                                                          & \citet{osten2006}                                                            &  & 46 & 59628.308    & NICER      & NICER Archive Website\textsuperscript{N}     \\
22 & 53469.053    & Arecibo                                                                          & \citet{osten2008}                                                           &  & 47 & 59628.447    & NICER      & NICER Archive Website\textsuperscript{N}     \\
23\tnote{h} & 57208.874    & VLA                                                                        &\citet{villadsen2019}                                                            &  & 48 & 59630.610    & NICER      & NICER Archive Website\textsuperscript{N}     \\
24\tnote{h} & 57222.836    & VLA      & \citet{villadsen2019}                               &  & 49 & 60151.354     & FAST      & This paper \\
25 & 57270.915   & VLA      & \citet{villadsen2019}      \\ \hline
\end{tabular}
  \begin{tablenotes}
  \item[h] Flares with (corresponding) X-ray band energy higher than 3$\times$10$^{33}$ erg, which we identify as high-energy flares.
 \item[N] https://heasarc.gsfc.nasa.gov/docs/nicer/nicer\_archive.html.
  \end{tablenotes} 
\end{threeparttable}
\label{dataset1}
\end{table*}

\begin{figure}
  \centering
  \includegraphics[width = 0.75\linewidth]{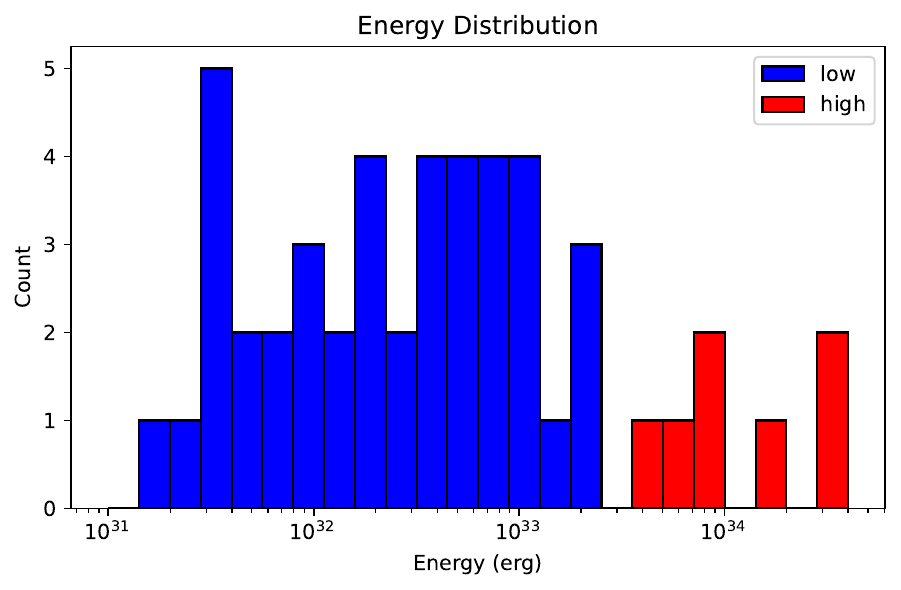}
\caption{The X-ray band energy distribution of AD Leo flares in dataset I. The blue histogram shows the energy distribution of 42 low-energy flares with their energy not exceeding 3$\times$10$^{33}$ erg. The red histogram shows the energy distribution of 7 high-energy flares.}
\label{fig:Energy}
\end{figure}

In order to check possible stellar rotation and/or SPI modulations of the bursts, we combine all detected radio and X-ray flares in the literature (dataset I) for periodicity analysis.
As shown in Table \ref{dataset1}, dataset I contains a total of 49 flares.
The burst energy of each flare is calculated.
The energy of X-ray flares is presented in the literature or can be calculated using the X-ray spectral-fitting program \texttt{XSPEC} \citep{arnaud1996}. 
For most radio flares, only the peak flux and time duration are specified in the literature.
So we multiply the FWHM of flare duration and the peak flux to estimate the burst energy.  
For flare 7, 10, 41, and 42, each composed of a number of sub-bursts, we adopt the duration from the start time of the first sub-burst to the end time of the last sub-burst as the flare duration.
The corresponding X-ray energy for radio flares is then calculated using the Güdel-Benz relation \citep{benz2010}.
It should be noted that the Güdel-Benz relation is based on the gyrosynchrotron emission. It may not be accurate for ECM emissions that are considered to be the major mechanism of the radio flares.
So the corresponding X-ray energy we calculated for radio flares is an order-of-magnitude estimation.
The distribution of X-ray energy for flares in dataset I is shown in Figure \ref{fig:Energy}. 
There are 7 flares with energy higher than $3\times10^{33}$ erg, which are considered as high-energy superflares \citep{schaefer2000}.

Flares in dataset I were detected by different telescopes at different wavebands. 
So the burst rates of detections, calculated by dividing the counts by the observation time, are not valid to compare to each other in the period analysis.
As flares detected by NICER have the largest population and the same detection threshold, we define them as dataset II. 
Then in the period analysis for dataset II, the count rate of bursts can be used.



\subsection{Method}


\begin{figure*}
  \centering
  \includegraphics[width = 0.6\textheight]{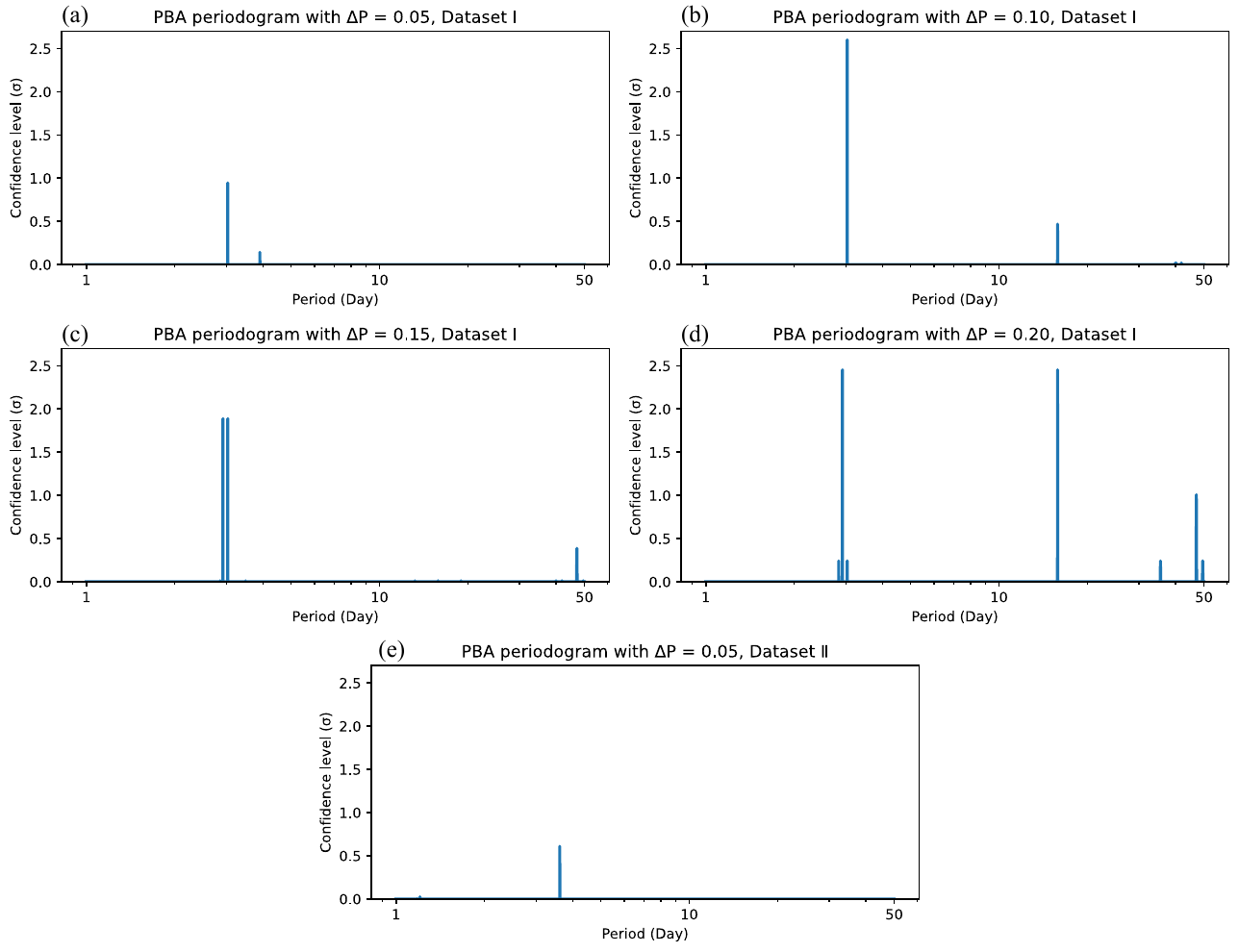}
\caption{The value of confidence level ($\sigma$) for test frequencies from $1$ to $50^{-1}$ day$^{-1}$  
(corresponds to test periods from $T = 1$ to $50$ days) calculated with different half-width $\Delta P$ for dataset I (a, b, c, and d) and dataset II (e).
Search results with a maximum confidence level higher than 0.5$\sigma$ are shown in this figure.
Maximum confidence levels of $2.60\sigma$ and $2.45\sigma$ are found for test periods $T = 3.04$ and $15.84$ day respectively, shown in panels (b) and (d).
The maximum confidence level found for dataset II is $0.61\sigma$ at test period $T = 3.62$ day with $P = 0.90$ and $\Delta P = 0.05$.
}
\label{fig:Heatmap}
\end{figure*}

Here we use PBA to find candidate periods of the flares. 
The method is proposed in \citet{gao2021} and developed when searching for periods in FRBs \citet{li2024}. 
We summarize the key processes of the method here for the readers' convenience.

The MJD of the burst event is first folded into the phase space for a test period $T$, i.e.
  \begin{equation}
  {\rm MJD}=CT+P'T,
  \label{equ:phase-folding}
  \end{equation}
where $C$ is an integer and $0\leq P'<1$. 
Then by introducing a test central active phase $P$ and the half-width of the active phase $\Delta P$, we calculate the number of flares in this active phase region $N(T, P, \Delta P)$.
Assuming a phase-independent flare distribution, the binomial probability that at least $N(T, P, \Delta P)$ of the total $N(T, P, 0.5)$ flares occur in the active region can be calculated by the cumulative distribution function of the binomial distribution, i.e.,
    \begin{eqnarray}
        F[N(T, P, \Delta P);N(T, P, 0.5), 2\Delta P]= \nonumber \\  
        \sum_{x=N(T, P, \Delta P)}^{N(T, P, 0.5)}B[x;N(T, P, 0.5), 2\Delta P],     
    \label{equ:F}
    \end{eqnarray}
where $B(x;N,2\Delta P)$ is the binomial probability of $x$ events to occur in $N$ experiments with probability for a single event being $2\Delta P$, i.e.,
\begin{eqnarray}
    &B[x;N(T, P, 0.5), 2\Delta P] = \nonumber \\
  &C_{N(T, P, 0.5)}^{x} (2\Delta P)^{x}(1- 2\Delta P)^{N(T, P, 0.5)-x}.
\label{equ:binomial probability}
\end{eqnarray}
From the cumulative probability $F$ of a single trial, one can determine the false alarm probability ($FAP$) follow the definition in \citet{horne1986} and \citet{vanderplas2018}, i.e.,
\begin{equation}
  FAP=1-(1-F)^{N_{\rm P}N_{\rm \Delta P}N_{\rm eff}}.
  \label{equ:FAP}
  \end{equation}
Here $N_{\rm P}$ and $N_{\rm \Delta P}$ represent the numbers of $P$ and $\Delta P$ searched in the trials.
The number of independent frequencies 
  \begin{equation}
  N_{\rm eff}=f_{\rm Ny}\Delta T,
  \label{equ:Neff}
  \end{equation}
where $\Delta T$ is the maximum time interval in the period search and the Nyquist frequency $f_{\rm Ny}$ is defined as
  \begin{equation}
  f_{\rm Ny}=\frac{0.5}{\delta _*},
  \end{equation}
where $\delta _*$ denotes the resolution of time measurements. 
In this work, we adopt the typical length of a burst ($\sim0.01$ day) as this resolution, i.e. $\delta _*=0.01$ day.

For dataset II, since all detections have the same energy threshold, period analysis can be carried out more accurately by accounting for the observation window function.
This means the probability for a single burst to occur in the active phase region is no longer $2\Delta P$ in Equation \ref{equ:F}.
It should be replaced by the normalized 
   observation time spent within the region of central active phase $P$ and half-width $\Delta P$.

If above calculation achieves a $FAP$ in close approximity to zero, the phase independence assumption of burst distribution is not valid.
This indicates an intrinsic accumulation of burst events at the active phase region for this test period.
We then calculate the confidence level of the $FAP$ to show how reliable the result is.

\subsection{Result}
\label{subsec:calculation result}

\subsubsection{PBA Results}

\begin{figure*}
\centering
  \includegraphics[width = 0.47\textheight]{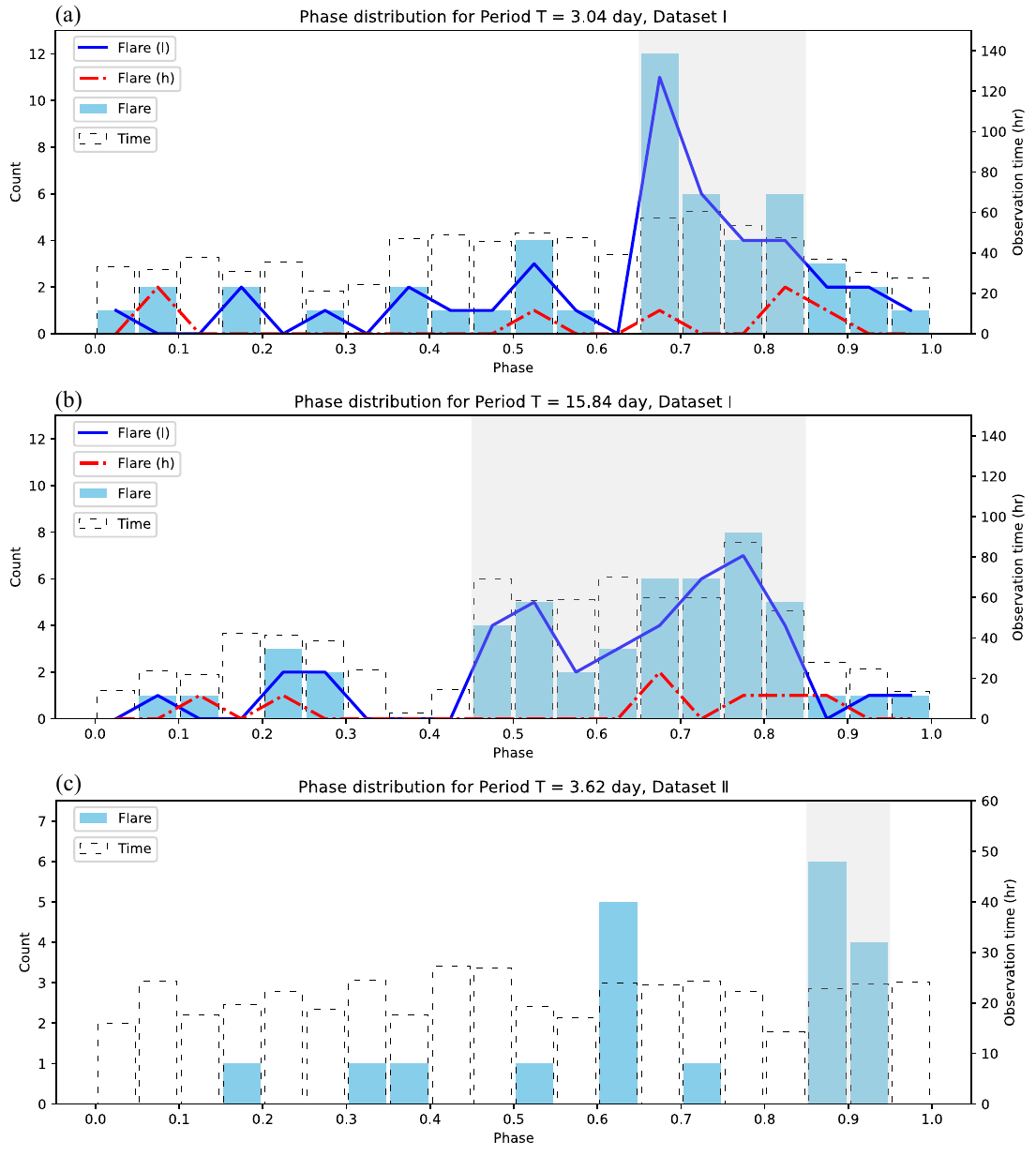}
\caption{Phase distributions for test periods $T = 3.04$ day (a) and $15.84$ day (b) on dataset I, and for test period $T = 3.62$ day (c) on dataset II.
The blue and dashed black histograms show flare counts and observation time respectively. The blue solid and red dash lines in panels (a) and (b) represent the phase distribution of low- and high-energy flares respectively. The active phase is the grey region.
Flares accumulate in the active region in all three panels.}
\label{fig:PD}
\end{figure*}

The range of period search can be set according to the time interval and sampling rate of the dataset and the candidate physical models of periodicity.
If a planet orbiting a star can lead to star-planet interaction, it is more likely a hot Jupiter enclosed in the magnetosphere of the star. 
For hot Jupiters, the situations of orbital periods shorter than 2.23 days, or equivalently orbital radii smaller than $\sim$0.025 AU are rare due to tidal disruptions \citep{trilling2002,dawson2018}. 
In addition, hot Jupiters usually have an orbital radius $<0.1$ AU \citep{udry2007}. 
So if a hot Jupiter is orbiting AD Leo, its orbital period is expected to be larger than the stellar spin period of 2.23 days, and smaller than 17.8 days (orbital period at 0.1 AU).
This corresponds to a beat period (of the planet's orbit and stellar spin) from 2.5 to $+\infty$ days if the orbit and spin are in the same direction.
When the planet's orbit and stellar spin are in opposite directions, which is not rare for hot Jupiters \citep{triaud2010}, the beat period can be 1.1 to 2.0 days accordingly. 
Additionally, for the dataset used here, the observation window is far from uniformly distributed in the phase space when the test period is larger than $\sim50$ day, possibly leading to pseudo-period detections.
So we finally set the test frequency to $1$ day$^{-1}$ to $50 ^{-1}$ day$^{-1}$, corresponding to test periods from $1$ day to $50$ day.
This covers most possible SPI periods (except for a phase-locked planet orbiting at approximately the stellar spin period and the spin direction), as well as the stellar spin period.
The test central active phase $P$ ranges from $0$ to $0.95$ with $0.05$ step, and the half-width $\Delta P$ from $0.05$ to $0.2$ with $0.05$ step.

The PBA periodogram is carried out for dataset I and II respectively.
The results with different active phase widths for dataset I, and with $\Delta P = 0.05$ for dataset II are shown in Figure \ref{fig:Heatmap}.
A higher confidence level indicates more likely the existence of a period for the flares.
There is no peak exceeding  3$\sigma$ confidence level.
Local maximums of 2$\sigma$ and 2.5$\sigma$ confidence levels at $T\approx3$ day and $T = 15.84$ day can be found in panels (b) and (d). 
For dataset II, the highest confidence level is 0.61$\sigma$ at $T = 3.62$ day.

There are variations of stellar magnetic fields during the large time intervals of  $15779$ days for dataset I and $1062$ days for dataset II \citep{bellotti2023}, which can erase flare periodicity.
Also, the large time intervals lead to large numbers of independent frequencies (Equation \ref{equ:Neff}).
So with the small number of flares, i.e., $49$ in dataset I and $20$ in dataset II, the false alarm probability $FAP$ in Equation 5 is usually not small enough to achieve a high confidence level. 


The phase distribution of flare counts and observation time for these three test periods are displayed in Figure \ref{fig:PD}.
Panel (b) in this figure shows that the flare events, as well as the observation time, are both concentrated in the same active region.
So the period of $T = 15.84$ day is likely an observation effect and consequently pseudo.
In contrast, observation time in panels (a) and (c) is distributed more uniformly.
So the concentrations of flares for $T = 3.04$ and $3.62$ day are not observation effects. 
Based on the $2.60\sigma$ confidence level, the period $T = 3.04$ day can be treated as a candidate period awaiting further tests.
The test period of $T = 3.62$ day, with a confidence level of only $0.61$, also shows concentrations of bursts to specific phases in Figure \ref{fig:PD}.
It is also noted in Figure \ref{fig:PD} that there is no significant difference in phase distributions for high- and low-energy flares.


\begin{figure}
\centering
  \includegraphics[width = 0.33\textheight]{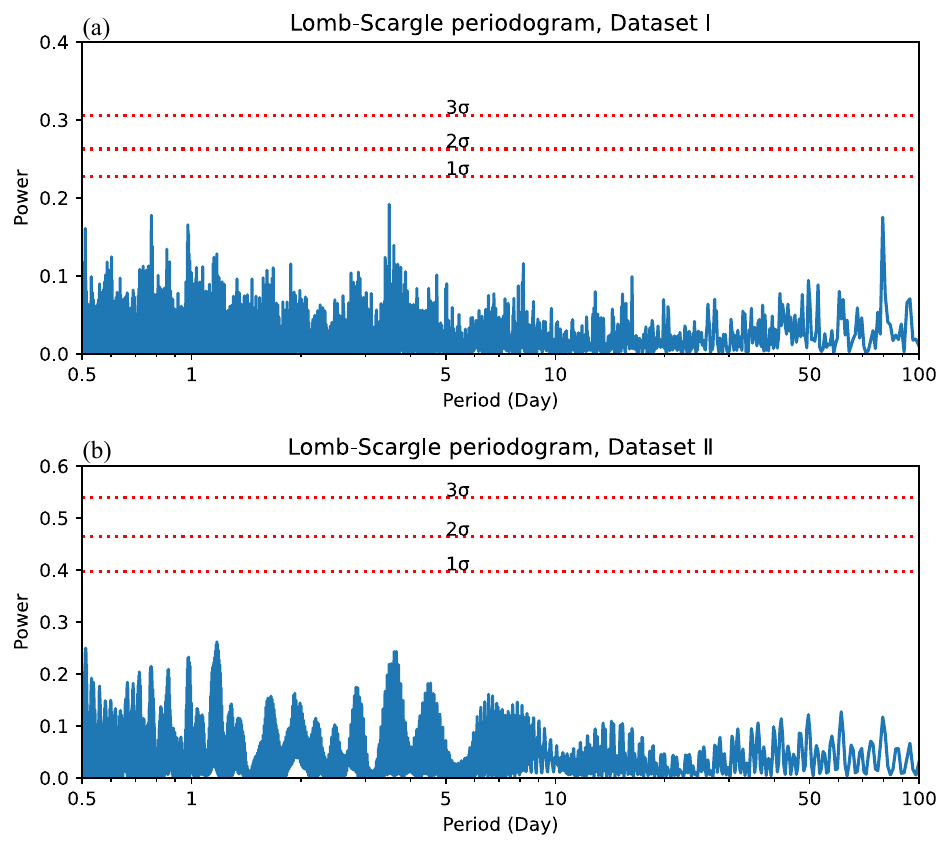}
\caption{Lomb-Scargle periodogram of dataset I (a) and II (b) with test frequency from $2$ to $100^{-1}$ day$^{-1}$ (corresponding to test period from 0.5 to 100 day).
The red dashed lines show the confidence levels of 1$\sigma$, 2$\sigma$, and 3$\sigma$ respectively.
The highest peak is lower than the 1-$\sigma$ level.
Local peaks at test period $T = 3.6$ day can be found in both panels (a) and (b).}
\label{fig:LS}
\end{figure}

\subsubsection{Lomb-Scargle Periodogram}

To check the results of the PBA method, we further adopt the Lomb-Scargle periodogram \citep{vanderplas2018} for dataset I and II respectively.
Following the method of periodicity analysis in \citet{cruces2021}, we first create a time series of 1 and 0 for bursts and non-detections from the observations of dataset I and II. 
The Lomb-Scargle periodogram is adopted for the time series, searching for frequencies from $2$ to $100^{-1}$ day$^{-1}$, corresponding to periods from 0.5 to 100 days. 
Results are shown in Figure \ref{fig:LS}, where the detection power is always below the 1$\sigma$ level.
However, it can be noted that there is a local peak at test period $T\approx3.6$ day in the Lomb-Scargle periodogram (Figure 6 (a) and (b)), consistent with the result of PBA (Figure \ref{fig:Heatmap} (e)).

While the current burst datasets did not yield a period detection, the candidate period at $\approx$3.0 days, as well as the notable test period at $\approx$3.6 days merit further investigation in future studies.
The comparison with the results of Lomb-Scargle periodogram, which is based on Fourier series, also indicates that PBA is more sensitive in detecting the accumulation of bursts in an active phase (cf. Figure \ref{fig:PD} (a)).

\section{Conclusions}
\label{sec:conclusion}
In this paper, we report the results of two observation campagins of AD Leo conducted in 2020 and 2023 using FAST. 
Based on the simultaneous increases in total flux density and circular polarization, a radio flare with $\sim$ 50\% left-hand circular polarization was identified on July 26, 2023. 
The flare may originate from ECM or other non-thermal radiations from the stellar magnetic field. 
The period analysis of AD Leo's radio and X-ray flares with PBA and Lomb-Scargle periodograms does not detect any periodicity at above 3$\sigma$ confidence level. 
But a candidate period at $\approx$3.0 day with $2.6\sigma$ confidence level is found in PBA; another period at $\approx3.6$ day shows local peaks in both PBA and Lomb-Scargle periodograms but at confidence levels lower than $1\sigma$. Both periods await further tests.


The FAST pulsar-end data and multi-waveband observations of the flare on July 26, 2023, will be presented in a follow-up work. 
The preliminary results of periodicity analysis suggest a more compact observation campaign on AD Leo.



\section*{Acknowledgements}
We acknowledge the anonymous referee for helpful discussions that improved this work. This work made use of the data from FAST (Five-hundred-meter Aperture Spherical radio Telescope). FAST is a Chinese national mega-science facility, operated by the National Astronomical Observatories, Chinese Academy of Sciences. 

This work is supported by the National SKA Program of China No. 2022SKA0120101, National Natural Science Foundation of China NSFC\# 42150105, and the Fundamental Research Funds for the Central Universities (Sun Yat-sen University, 2021qntd28, 22qntd3101).  

Lei Qian is supported by the National SKA Program of China No. 2020SKA0120100, the National Nature Science Foundation of China NSFC\# 12173053, the Youth Innovation Promotion Association of CAS (id.~2018075), and the CAS ``Light of West China" Program.

\section*{Data Availability}
The data underlying this article will be shared on reasonable request to the corresponding author.





\bibliographystyle{mnras}
\bibliography{mylab_0517}




\appendix

\section{Additional information on data processing}

\subsection{Polarization calibration}
\label{appendix:polarization calibration}
\begin{figure}
    \centering
    \includegraphics[width = 0.7\linewidth]{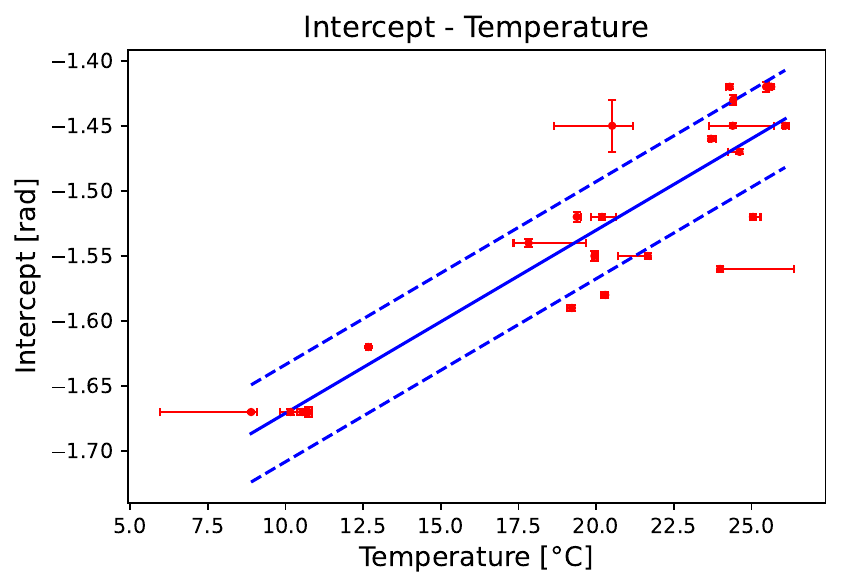}
    \caption{The correlation between the real-time environmental temperature and the intercept in the $\Delta\delta_{\rm ref}$ - frequency relation. The dashed lines are the range of the error estimated by the root mean square.}
    \label{fitting}
\end{figure}

As the noise diode data in 2020 AD Leo observations cannot be extracted due to the large sampling time, 
we adopt FAST observations conducted during the same 17 days with the same observation mode and backend settings to carry out polarization calibration. 
$\Delta (AA_{\rm ref} / BB_{\rm ref})$ in these observations has a mean of $0.12\pm0.16$, so we adopt it as 0.12 in the calibration.
On the other hand, the value of $\Delta\delta_{\rm ref}$ has a linear relation with frequency.
The slope in the linear relation is a constant, and the intercept varies for different observations. 
This variation arises from the different environmental temperatures when the observations are carried out. 
There is again a linear correlation between the intercept and the environmental temperature of the observation, as shown in Figure \ref{fitting}. 
Hence, with the environmental temperature recorded every second, we obtain the real-time intercept values in observations of AD Leo.
Consequently, the phase delays $\Delta\delta_{\rm sig}$ are calculated at different frequencies to correct the observation signals.

\subsection{Removal of RFI and standing waves}
\label{appendix:removal}
\begin{figure}
    \centering
    \includegraphics[width = \linewidth]{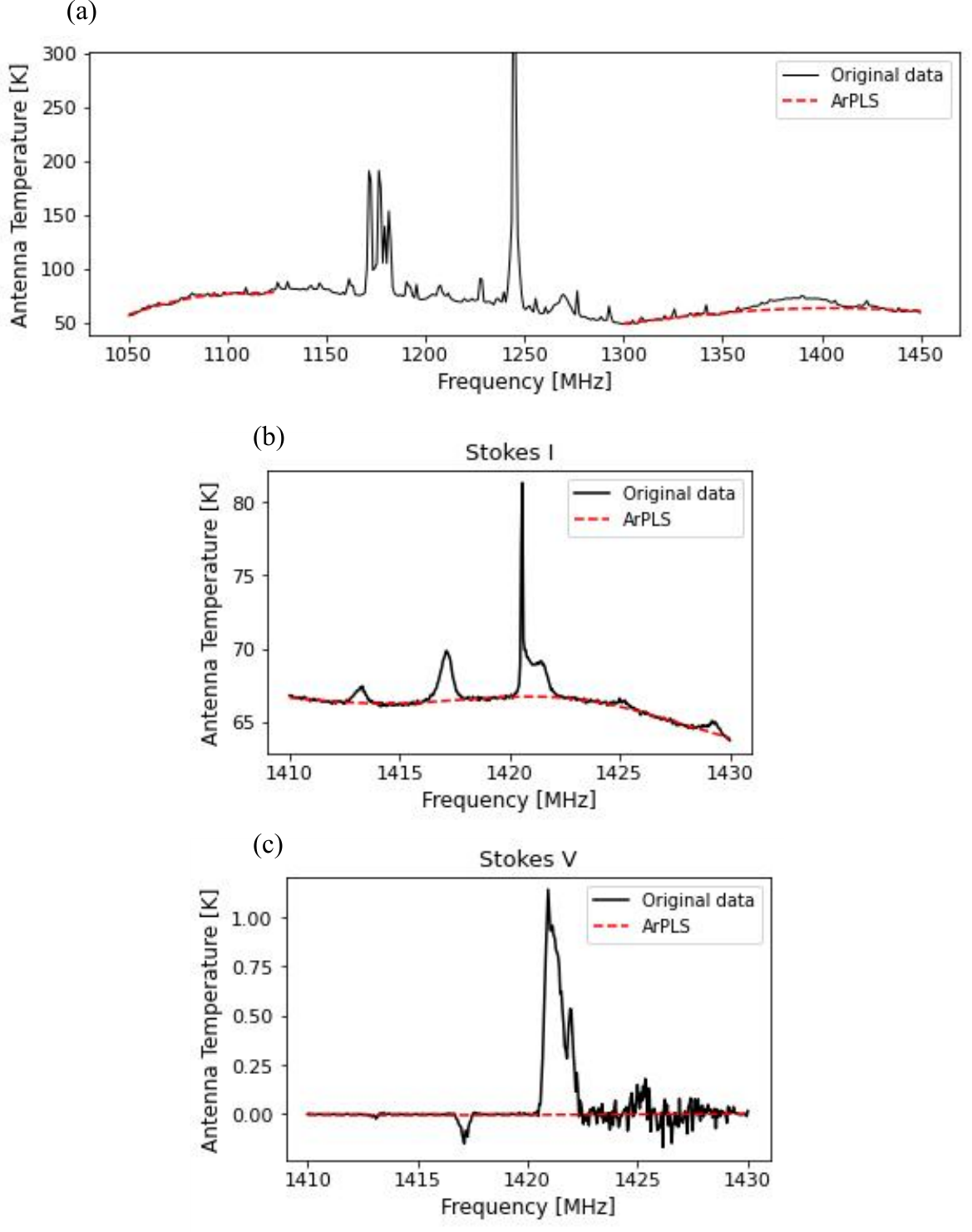}
    \caption{Illustration of how the continuum is achieved using ArPLS for the spectrum of the full waveband in Stokes I (a), of the HI protected waveband in Stokes I (b) and Stokes V (c) for the November 8 observation in 2020.}
    \label{arpls}
\end{figure}

The RFI and standing waves are additives to the continuum. We apply the ArPLS algorithm \citep{baek2015} to remove them and get the continuum used in this paper. Basically, this algorithm is a least square smoothing method. By changing the weights of signals iteratively, RFIs, HI lines, as well as standing waves are removed sequentially. The performance of ArPLS in this work is depicted in Figure \ref{arpls} for the observation on November 8, 2020, as an example.

\bsp	
\label{lastpage}
\end{document}